\documentclass[prd,twocolumn,showpacs,superscriptaddress,nofootinbib,floatfix,showkeys,reprint]{revtex4-2}
\usepackage{graphicx}
\usepackage{amsmath}
\usepackage{bm}
\usepackage{yhmath}
\usepackage{mathtools}
\usepackage{wasysym}
\usepackage[colorlinks,citecolor=blue,urlcolor=blue,linkcolor=blue]{hyperref}
\usepackage{subfigure}
\usepackage{color}
\usepackage{cases}
\usepackage{subfigure}
\usepackage{times}
\usepackage{dcolumn,booktabs,bm}
\usepackage{slashed}
\usepackage{amsfonts,amssymb,stmaryrd,latexsym,amsmath}
\usepackage{textcomp}
\usepackage{multirow}
\usepackage{cancel}
\usepackage{array}

\def\SU3{{\text{SU(3)}_{\rm F}}}
\def\DD{{\bar{D}_{(s)}^{(*)}D_{(s)}^{(*)}}}
\def\tP{{\mathtt{P}}}
\def\tV{{\mathtt{V}}}

\renewcommand{\arraystretch}{1.8}

\begin{document}

\title{Implications of the $Z_{cs}(3985)$ and $Z_{cs}(4000)$ as two different states}

\author{Lu Meng}
\affiliation{Institut f\"ur Theoretische Physik II, Ruhr-Universit\"at Bochum,  D-44780 Bochum,	Germany}

\author{Bo Wang}
\affiliation{School of Physical Science and Technology, Hebei
University, Baoding 071002, China}

\author{Guang-Juan Wang}\email{wgj@pku.edu.cn}
\affiliation{Advanced Science Research Center, Japan Atomic Energy
	Agency, Tokai, Ibaraki, 319-1195, Japan}

\author{Shi-Lin Zhu}\email{zhusl@pku.edu.cn}
\affiliation{School of Physics and Center of High Energy Physics,
Peking University, Beijing 100871, China}

\begin{abstract}
Recently, the hidden charm tetraquark states $Z_{cs}(3985)$ and $Z_{cs}(4000)$ with strangeness were observed by the BESIII and LHCb collaborations, respectively, which are great breakthroughs for exploring exotic QCD structures. The first and foremost question is whether they are the same state. In this work, we explore the implications of the narrower state
$Z_{cs}(3985)$ in BESIII and the wider one $Z_{cs}(4000)$ in LHCb
as two different states. Within a solvable nonrelativistic
effective field theory,  we include the possible violations of heavy
quark spin symmetry and SU(3) flavor symmetry in a comprehensive approach. If $Z_{cs}(3985)$ and $Z_{cs}(4000)$ are two different states, our results show that $Z_{cs}(4000)/Z_{cs}(3985)$ is the pure
$(|\bar{D}_s^*D\rangle+/- |\bar{D}_sD^*\rangle)/\sqrt{2}$ state, and the SU(3) flavor partner of $Z_{c}(3900)$ is $Z_{cs}(4000)$ rather than the $Z_{cs}(3985)$. Another two important consequences are the existence of a tensor $\bar{D}_s^*D^*$ resonance with mass about 4126 MeV and width 13 MeV, and the suppression of the decay mode $Z_{cs}(3985) \to J/\psi K$. The two consequences can be tested in experiments and distinguish the two-state interpretation from the one-state scheme.
\end{abstract}

\maketitle

\thispagestyle{empty}

\section{Introduction}
The quark model provides a very successful classification scheme for the conventional hadrons in terms of the valence quarks,  the mesons ($\bar q q$) and baryons ($qqq$). It has been experimentally verified during the past time. However, in 2003, the observation of the $X(3872)$ \cite{Choi:2003ue}, which does not fit into the quark model makes the hadron spectroscopy become again a challenging  problem. Since then, the pace to exploring the QCD exotica beyond the simple quark model, the so-called  XYZ exotic states, has never stopped~\cite{Chen:2016qju,Guo:2017jvc,Liu:2019zoy,Lebed:2016hpi,Hosaka:2016pey,Olsen:2017bmm,Karliner:2017qhf,Brambilla:2019esw,Esposito:2016noz}. The observations of the $Z_c$ states~\cite{Liu:2013dau,Ablikim:2013mio,Ablikim:2013wzq} and the $P_c$ states~\cite{Aaij:2015tga,Aaij:2019vzc} are the milestones in the exploration of the exotic states. According to their charges  and decay channels, their minimum quark constituents are  $c\bar{c}q\bar{q}$ and $c\bar{c}qqq$ rather than the $c\bar c$ in the conventional charmonium. This  makes them to be the multiquark states undoubtedly. 

Recently, the LHCb and BESIII collaborations made great
breakthroughs in searching for hidden charmed multiquark states with
strangeness~\cite{Aaij:2020gdg,Ablikim:2020hsk,Aaij:2021ivw}, which starts a new epoch of exotic hadrons.
Before these
experimental observations, the hidden charm tetra- and
pentaquarks with strangeness were predicted~\cite{Wu:2010jy,Wang:2019nvm,Ferretti:2020ewe}. LHCb collaboration obtained the first evidence of the pentaquark
state with strangeness~\cite{Aaij:2020gdg}. BESIII
collaboration reported a structure $Z_{cs}(3985)^-$ in the $K^+$
recoil-mass spectra in the $e^+e^-\to K^+(D_s^-D^{*0}+D_s^{*-}D^0)$
with $5.3$ $\sigma$~\cite{Ablikim:2020hsk}. Several months later,
LHCb collaboration announced the observation of two strange hidden
charm states, $Z_{cs}(4000)^+$ and $Z_{cs}(4220)^+$ in the invariant
mass spectrum of the $J/ \psi K^+$ channel in the $B^+\to J/\psi
\phi K^+$ process with the significance of 15 $\sigma$ and 5.9
$\sigma$, respectively~\cite{Aaij:2021ivw}.

\begin{figure*}[!htp]
	\centering
	\includegraphics[width=0.99\textwidth]{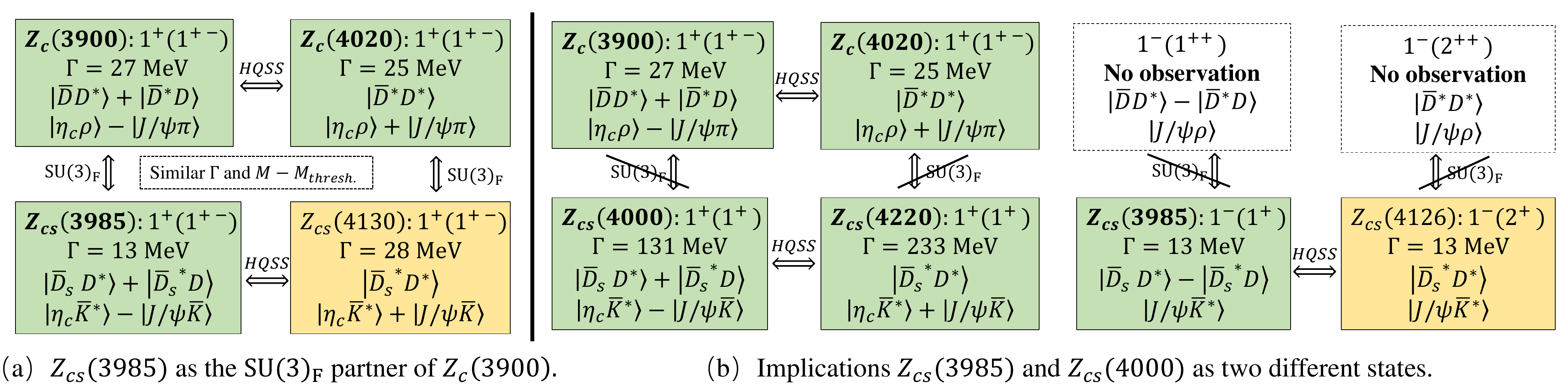}
	\caption{ Comparison of two assignments of $Z_{cs}(3985)$ and $Z_{cs}(4000)$ states. The information in subfigure (a) is extracted from from Refs.~\cite{Meng:2020ihj,Wang:2020htx} and subfigure (b) illustrates the main conclusions of this work treating the two $Z_c$s as  different states. The particles in green and yellow cards are observations and our predictions, respectively.  The quantum numbers $I^{G_I}/U^{G_U}/V^{G_V}(J^{PC})$ of the states are given in the first row of each card. $\Gamma$ is the decay width. The hadronic components are listed in the third row of each card. In the last row, the wave function on the basis of  the heavy quark symmetry $|({c\bar{c}})_{S_H}({q_1\bar{q}_2})_{S_L}\rangle$ (${S_H}$ and $S_L$ are the heavy and light degrees of freedom, respectively) is given after quark rearrangement according to Eqs.~\eqref{eq:rearange1}-\eqref{eq:rearange4}. We omit the possible $1/\sqrt{2}$ factor for conciseness. The corresponding ground charmonium and the light mesons in the kets represent the $S_H$ and $S_L$ in the last row, representively. They are the possible hidden charmed decay modes of the $Z_c/Z_{cs}$ states if allowed by the phase space in the heavy quark symmetry.  We use the double head arrows to denote partner states in HQSS and $\SU3$ symmetry. In subfigure (b), we use the oblique lines to represent the breaking of $\SU3$ symmetry. For completeness, we list the information of two partner channels in white cards, which are neither the experimental resonances nor our theoretical predictions.}\label{fig:twocase}
\end{figure*}

The observation of $Z_{cs}(3985)^-$ in BESIII inspired amounts of
theoretical works to explore its
nature~\cite{Meng:2020ihj,Yang:2020nrt,Cao:2020cfx,Du:2020vwb,Wang:2020htx,Wang:2020rcx,Sun:2020hjw,Chen:2020yvq,Wang:2020kej,Azizi:2020zyq,Jin:2020yjn,Wan:2020oxt,Shen:2020gpw,Ikeno:2021ptx,Yan:2021tcp,Ozdem:2021yvo,Albuquerque:2021tqd,Xu:2020evn,Simonov:2020ozp,Wang:2020iqt,Wang:2020dmv},
which had been released before the observation of $Z_{cs}(4000)$ and
$Z_{cs}(4220)$ in LHCb collaboration. Since the $Z_{cs}(3985)$ lies close to the mass threshold of the $\bar{D}_sD^*/\bar{D}_s^*D $ channel, a natural and popular
interpretation of the $Z_{cs}(3985)$  is the loosely bound molecule composed of the two mesons,
$\bar{D}_sD^*/\bar{D}_s^*D $, which is the SU(3)
flavor [$\SU3$] symmetry partner of the
$Z_{c}(3900)$~\cite{Liu:2013dau,Ablikim:2013mio}. In Refs.~\cite{Meng:2020ihj,Wang:2020htx}, the hadronic components of $Z_c(3900)$ and $Z_{cs}(3985)$ are given
 \begin{equation}
	|\bar{\mathtt{P}}\mathtt{V}/\bar{\mathtt{V}}\mathtt{P},G_{I/U/V}\rangle =\frac{1}{\sqrt{2}}\left(|\bar{\mathtt{P}}\mathtt{V}\rangle+G_{I/U/V}|\bar{\mathtt{V}}\mathtt{P}\rangle\right),\label{eq:G_parity}
\end{equation}
where $\mathtt{P}$ and $\mathtt{V}$ are the pseudoscalar and vector
heavy-light mesons, respectively. $G_{I/U/V}=\pm1$ is the generalized $G$-parity, where conventional parity is extended to $U$-spin and $V$-spin sectors. $G_{I/U/V}$ of $Z_c(3900)$ and $Z_{3985}$ are both $+1$ in Refs.~\cite{Meng:2020ihj,Wang:2020htx}. The $Z_{cs}(3985)$ was assigned
 as the partner state of $Z_{c}(3900)$ due to their
consistent masses with respect to thresholds, widths and line
shapes. A natural
prediction is a vector $\bar{D}^*_sD^*$ di-meson state as the heavy
quark spin symmetry (HQSS) partner of the $Z_{cs}(3985)$ and the
$\SU3$ partner of the $Z_c(4020)$~\cite{Ablikim:2013wzq}. The
typical features or assumptions of above interpretations are the
good HQSS and $\SU3$ symmetry as shown in Fig.~\ref{fig:twocase}~(a).

\begin{figure}[!htp]
	\centering  \includegraphics[width=0.35\textwidth]{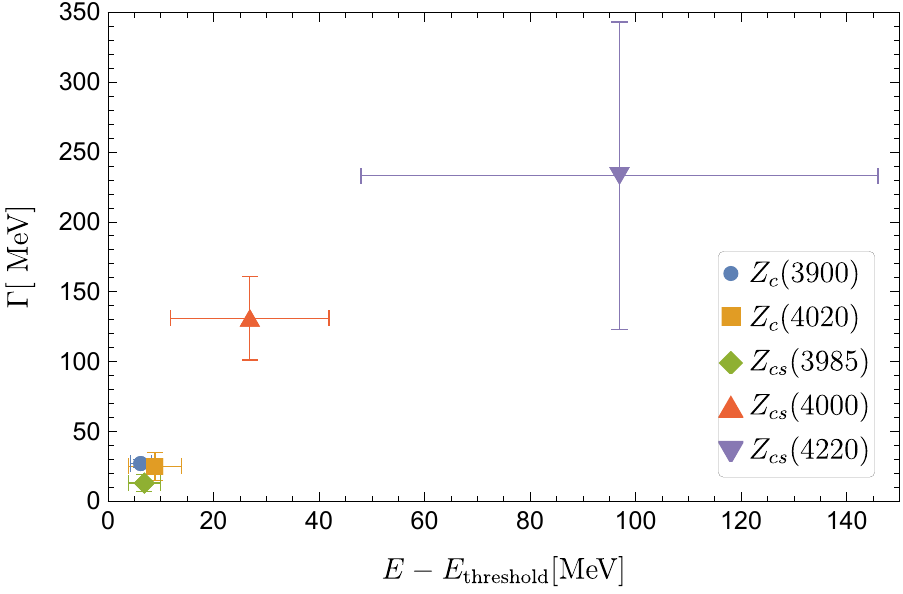}
	\caption{  Masses with respect to relevant thresholds and decay widths of $Z_c$ and $Z_{cs}$ states~\cite{Ablikim:2020hsk,Aaij:2021ivw,Ablikim:2013emm,Ablikim:2015swa}. The corresponding thresholds for $Z_c(3900)$, $Z_c(4020)$, $Z_{cs}(3985)$, $Z_{cs}(4000)$ and $Z_{cs}(4220)$ are $\bar{D}D^*/\bar{D}^*D$,  $\bar{D}^*D^*$, $\bar{D}_{s}D^*/\bar{D}_{s}^*D$, $\bar{D}^*D_{s}/\bar{D}D_{s}^*$ and $\bar{D}^*D_{s}^*$, respectively.}\label{fig:mss_wd}
\end{figure}

The $Z_{cs}(4000)$ was observed in LHCb in the proximity of the
$\bar{D}_sD^*/\bar{D}_s^*D $ thresholds like the $Z_{cs}(3985)$.
However, the width of $Z_{cs}(4000)$ is over 100 MeV, which is about
ten times larger than that of $Z_{cs}(3985)$ as shown in
Fig.~\ref{fig:mss_wd}. Moreover, the
$Z_{cs}(4220)$ state with higher mass is also a much broader
resonance than the theoretical predictions of the HQSS partner of
$Z_{cs}(3985)$~\cite{Meng:2020ihj,Yang:2020nrt,Cao:2020cfx,Du:2020vwb,Wang:2020htx}.
Thus, theorists resorted to considering the $Z_{cs}(4000)$ and
$Z_{cs}(3985)$ as two different states in the hadronic molecular~\cite{Chen:2021erj} and the compact tetraquark
schemes~\cite{Maiani:2021tri}. The observations in LHCb also
triggered a new round of  theoretical investigations on the $Z_{cs}$
states~\cite{Liu:2021xje,Ortega:2021enc,Chen:2021uou,Ge:2021sdq}.

Up to now, most theoretical works only focused on the $Z_{cs}(3985)$ states, or neglected the large width difference between the  $Z_{cs}(3985)$ and $Z_{cs}(4000)$ and treated them as the same state. In fact, the significant difference between $Z_{cs}(4000)$ and $Z_{cs}(3985)$ were not taken seriously. In this work, we will explore the possibility of $Z_{cs}(3985)$ and
$Z_{cs}(4000)$ as two different di-meson resonances. We will
introduce the violation effects of HQSS and $\SU3$ symmetry in the
framework of a solvable nonrelativistic effective field theory
(NREFT).  Our numerical results and phenomenological analysis will
provide several important consequences of $Z_{cs}(3985)$ and
$Z_{cs}(4000)$ as two different states, including their hadronic
components, selection rules of the decay and the existence of a
tensor $\bar{D}_s^*{D}^*$ di-meson state. We will show that searching for the tensor
$\bar{D}_s^*D^*$ state and the $Z_{cs}(3985) \to J/\psi K$ decay in experiments can help us scrutinize the inner configurations of
$Z_{cs}(3985)$ and $Z_{cs}(4000)$.

\section{Symmetries and their violations}
The Hamiltonian of the $\DD$ di-meson systems can be divided into
the free part $\hat{H}_0$ and the interacting part $\hat{V}$. The $\SU3$ symmetry and HQSS
are both approximate symmetries of the Hamiltonian. For the
near-threshold states, the interaction is weak.   In
Ref.~\cite{Meng:2020cbk}, It was shown that the leading violations
of the two symmetries arise from the mass terms in $\hat{H}_0$. There are two large mass splittings compared to
the interaction $\hat{V}$,
 \begin{eqnarray}
    m_{D_{s}^{(*)}}-m_{D^{(*)}}&\simeq100\text{ MeV},\label{eq:mssu3}
    \\m_{D_{(s)}^{*}}-m_{D_{(s)}}&\simeq140\text{ MeV},\label{eq:masshq}
\end{eqnarray}
which will break the $\SU3$ symmetry and HQSS, respectively. In
this work we will consider the symmetry breaking effects from both the mass term and interaction.

In the spin space, the eigenvectors of $\hat{H}_0$ are $|\bar{\tP}\tP\rangle$ ($J^{P}=0^+$), $|\bar{\tP}\tV/\bar{\tV}\tP\rangle$  ($J^{P}=1^+$), and $|\bar{\tV}\tV \rangle$ ($J^{P}=0^+/1^+/2^+$) states. The mixing effect among them with the same quantum number would be suppressed by the mass splittings in Eq.~\eqref{eq:masshq}. The $|\bar{\tP}\tV\rangle$ and $|\bar{\tV}\tP\rangle$ states are (almost) degenerate for the free Hamiltonian and will mix with each other (The mass splitting between the $D_s^-D^{*0}$ and $D_s^{*-}D^0$ channels is only 2 MeV). The mixing angle will be determined by the interaction $\hat{V}$.

The interaction in the spin space could be introduced
as~\cite{Meng:2019ilv,Wang:2019ato},
\begin{eqnarray}
&V^{s}  =V_{q\bar{q}}^{s}+\text{HQSS breaking terms},\\
&V_{q\bar{q}}^{s}=c_{1}^{s}+c_{2}^{s}\bm{s}_{q}\cdot\bm{s}_{\bar{q}},
\end{eqnarray}
where $V_{q\bar{q}}^{s}$ is the interaction between the light
degrees of freedom, which satisfies the HQSS. The
$\bm{s}_{q/\bar{q}}$ is the spin operator of the light (anti)-quark.
$c_1^s$ and $c_2^s$ are two independent coupling constants. 

For the  $|\bar{\tP}\tV/\bar{\tV}\tP\rangle$  states, the hadronic
interactions in the
$\{\bar{\mathtt{P}}\mathtt{V},\bar{\mathtt{V}}\mathtt{P}\}$ basis
read,
\begin{eqnarray}
\langle
V^{s}\rangle_{\{\bar{\mathtt{P}}\mathtt{V},\bar{\mathtt{V}}\mathtt{P}\}}^{1^{+}}=\left[\begin{array}{cc}
    c_{1}^{s} & -\frac{1}{4}c_{2}^{s}\\
    -\frac{1}{4}c_{2}^{s} & c_{1}^{s}+\delta c^{s}
\end{array}\right],\label{eq:vs_pvvp}
\end{eqnarray}
where we introduce the $\delta c^s$ in the diagonal term to account for the HQSS violation effect. In the HQSS limit ($\delta c^s =0$), the interactions in the
$|\bar{\tP}\tV/\bar{\tV}\tP\rangle$ channels are the same, and the
eigenvectors of the Hamiltonian read,
\begin{eqnarray}
|\bar{\mathtt{P}}\mathtt{V}/\bar{\mathtt{V}}\mathtt{P},\pm\rangle &=&\frac{1}{\sqrt{2}}\left(|\bar{\mathtt{P}}\mathtt{V}\rangle\pm|\bar{\mathtt{V}}\mathtt{P}\rangle\right),
\end{eqnarray}
where the relative signs of two hadronic components are used to
label the states. For the $[\bar{D}D^*/\bar{D}^*D]^{I=1}$ and
$\bar{D}_sD^*/\bar{D}_s^{*}D$ states, the sign is that of the
$G_{I/U/V}$-parity as shown in Eq.~\eqref{eq:G_parity} (see
Ref.~\cite{Meng:2020ihj} for details).

Another two related
channels are the $|\bar{\tV}\tV,1^+ \rangle$ and $|\bar{\tV}\tV,2^+
\rangle$. Their interactions in the HQSS limit read,
\begin{eqnarray}
        \langle V_{q\bar{q}}^{s}\rangle_{\{\bar{\mathtt{V}}\mathtt{V}\}}^{1^{+}}    &=& \langle V_{q\bar{q}}^{s}\rangle_{\{\bar{\mathtt{P}}\mathtt{V}/\bar{\tV}\tP,+\}}^{1^{+}}=c_{1}^{s}-\frac{1}{4}c_{2}^{s},\label{eq:hqss_vector}\\
    \langle V_{q\bar{q}}^{s}\rangle_{\{\bar{\mathtt{V}}\mathtt{V}\}}^{2^{+}}    &=& \langle V_{q\bar{q}}^{s}\rangle_{\{\bar{\mathtt{P}}\mathtt{V}/\bar{\tV}\tP,-\}}^{1^{+}}=c_{1}^{s}+\frac{1}{4}c_{2}^{s}.\label{eq:hqss_tensor}
\end{eqnarray}
The two equalities indicate the existence of HQSS partners~\cite{Nieves:2012tt,Baru:2016iwj}. For
example, $Z_c(3900)$ and $Z_c(4020)$ have similar masses with
respect to the corresponding thresholds and decay widths, which
stems from the Eq.~\eqref{eq:hqss_vector}. The existence of $Z_c(3900)$ and
$Z_c(4020)$ as the partner states indicates that the HQSS is a good
approximation. It is reasonable to infer a similar property for the
$\bar{D}_s^{(*)}D^{(*)}$ systems. We will prove this
conclusion  numerically and predict a tensor
$\bar{D}_s^*D^*$ state with the $Z_{cs}(3985)$ and $Z_{cs}(4000)$ as two
different states.

We can rewrite the above four states on the basis of the heavy quark
symmetry $|S^P_{\bar{c}c},S^P_{q_1\bar{q}_2};J^P\rangle$,
\begin{eqnarray}
    &&|\bar{\mathtt{P}}\mathtt{V}/\bar{\mathtt{V}}\mathtt{P},+\rangle   =\frac{1}{\sqrt{2}}\left(|0_{\bar{c}c}^{-},1_{q_{1}\bar{q}_{2}}^{-};1^{+}\rangle-|1_{\bar{c}c}^{-},0_{q_{1}\bar{q}_{2}}^{-};1^{+}\rangle\right),\label{eq:rearange1}\\
    &&|\bar{\mathtt{P}}\mathtt{V}/\bar{\mathtt{V}}\mathtt{P},-\rangle   =|1_{\bar{c}c}^{-},1_{q_{1}\bar{q}_{2}}^{-};1^{+}\rangle,\\
    &&|\bar{\mathtt{V}}\mathtt{V},1^{+}\rangle=\frac{1}{\sqrt{2}}\left(|0_{\bar{c}c}^{-},1_{q_{1}\bar{q}_{2}}^{-};1^{+}\rangle+|1_{\bar{c}c}^{-},0_{q_{1}\bar{q}_{2}}^{-};1^{+}\rangle\right),\\
    &&|\bar{\mathtt{V}}\mathtt{V},2^+\rangle=|1_{\bar{c}c}^{-},1_{q_{1}\bar{q}_{2}}^{-};2^{+}\rangle.\label{eq:rearange4}
\end{eqnarray}
The wave function decomposition would give rise to a series of
selection rules for the decay modes of the
$\bar{D}_{(s)}^{(*)}D^{(*)}$ di-meson systems.

In the flavor space, the leading $\SU3$ breaking effect
stems from the mass splitting in Eq.~\eqref{eq:mssu3}, which will
suppress the mixing effect between
$(|u\bar{u}\rangle+|d\bar{d}\rangle)/\sqrt{2}$ and
$|s\bar{s}\rangle$ systems (We omit the heavy quark part for conciseness .)~\cite{Meng:2020cbk}. For the isovector
and open strange systems, the $\SU3$ violation effect starts to
appear at the interaction terms. Unlike Ref.~\cite{Meng:2020ihj}, we neither presume the $\SU3$
symmetry is a good approximation nor relate the interaction of
$[\bar{D}^{(*)}D^{(*)}]^{I=1}$ to those of $\bar{D}^{(*)}_sD^{(*)}$.
We will show that the $Z_{cs}(3985)$ and $Z_{cs}(4000)$ as two
different states imply the large $\SU3$ violation effect.

\section{Solvable NREFT for resonances}
We adopt a solvable NREFT to investigate the $Z_c$ and $Z_{cs}$
states as the $S$-wave di-meson resonances~\cite{Hammer:2019poc}. We construct the general
contact $S$-wave interaction to the next-to-leading order,
\begin{eqnarray}
    V(\bm{p},\bm{p}')   =\frac{c_{a}}{\Lambda}+\frac{c_{b}}{\Lambda^{3}}(\bm{p}^{2}+\bm{p}'^{2}),\label{eq:ptl_single}
\end{eqnarray}
where $c_a$ and $c_b$ are the low energy constants (LECs) for the
leading order (LO) and the next-to-leading order (NLO). $\Lambda$ is
the cutoff scale. The possible $\bm{p}\cdot \bm{p'}$ term becomes
vanishing after the partial wave expansion for the $S$-wave channel.
Similar interaction was adopted in
Refs.~\cite{Albaladejo:2015lob,Meng:2020ihj}.

We first adopt the single-channel formalism. The $Z_c(Z_{cs})$ states are molecule-type states barely above the thresholds. The fall-apart decaying into their constituent hadrons would be dominant. $Z_{c}(Z_{cs})$ couples to the  $J/\psi \pi(K) $ channel  through recluster of the heavy quarks, which is very weak. In Ref.~\cite{Meng:2020ihj}, a coupled-channel calculation considering the  $J/\psi \pi (J/\psi K)$ also shown that the such channels play minor roles, which neither changes the property of the pole nor contributes to the large partial decay with.
Therefore, we only include the open-charmed
$\bar{D}^{(*)}_{(s)}D^{(*)}$ channels. The $\bar{\tP}\tV/\bar{\tV}\tP$ and $\bar{\tV}\tV$ thresholds are different by about 100MeV as shown in Eq.~\eqref{eq:masshq}, which would suppressed their mixing effect. Therefore, we deal with
$\bar{\tP}\tV/\bar{\tV}\tP$ and $\bar{\tV}\tV$ channels separately.  For the $Z_c(3900)$,
$Z_c(4020)$, $Z_{cs}(3985)$ and $Z_{cs}(4000)$ states, we adopt four
sets of independent LECs in Eq.~\eqref{eq:ptl_single}, which will be
determined by the mass and width of the corresponding state.

The single-channel Lippmann-Schwinger equation
(LSE) reads,
\begin{eqnarray}
    T(\bm{p},\bm{p}')=V(\bm{p},\bm{p}')+\int\frac{d^{3}\bm{q}}{(2\pi)^{3}}\frac{V(\bm{p},\bm{q})T(\bm{q},\bm{p}')}{E-\frac{\bm{q}^{2}}{2\mu}+i\epsilon},\label{eq:ptl}
\end{eqnarray}
where $E$ is the energy with respect to the mass threshold and $\mu$ is
the reduced mass of the two heavy meson. The resonance corresponds to a pole of the $T$-matrix in the nonphysical  Riemann sheet.  A general numerical approach to solving the LSEs is the matrix-inversion method.

For the
specific separable interaction in Eq. \eqref{eq:ptl_single}, the LSEs can be solved analytically~\cite{Epelbaum:2017byx}. For the single-channel LSE,
the inverse of the $T$-matrix reads,
\begin{eqnarray}\label{eq:Tmatrix}
{1\over T(k)}=\frac{-G_{0}\Lambda^{5}c_{a}-2G_{2}\Lambda^{3}c_{b}+\left(G_{2}^{2}-G_{0}G_{4}\right)c_{b}^{2}+\Lambda^{6}}{\Lambda^{5}c_{a}+c_{b}\left[c_{b}\left(G_{0}k^{4}-2G_{2}k^{2}+G_{4}\right)+2k^{2}\Lambda^{3}\right]},\nonumber\\
\label{eq:inveseT}
\end{eqnarray}
where $k\equiv\sqrt{2\mu E}$ and $G_n(k)$ is defined as
\begin{eqnarray}
G_{n}=\int\frac{d^{3}\bm{q}}{(2\pi)^{3}}\frac{q^{n}}{E-\frac{q^{2}}{2\mu}+i\epsilon}.
\end{eqnarray}
Here, we  adopt the cutoff regularization ($\Lambda$ is
the cutoff parameter).  In the cutoff regularization scheme, we obtain the $G_n$,
\begin{eqnarray}
G_{0}&=&\frac{4\pi}{(2\pi)^{3}}2\mu\left[k\tanh^{-1}\left(\frac{k}{\Lambda}\right)-\Lambda-i\frac{\pi}{2}k\right],\label{eq:G0}\\
G_{n}
&=&k^{2}G_{n-2}-\frac{\mu}{\pi^{2}}\frac{\Lambda^{n+1}}{n+1}.
\end{eqnarray}
The recursive relation relates the $G_n$ to $G_{n-2}$. In our calculation, only the even numbered $n$ cases are involved in. Thus, we only give the $G_0$ explicitly here and obtain the $G_2$ and $G_4$ in Eq. \eqref{eq:Tmatrix} by the recursive relation.

There are two unknown LECs $c_a$ and $c_b$ in each $D^{(*)}_{(s)}\bar D^{(*)}$ channel. We treat $Z_c$ and $Z_{cs}$ states as resonances and their poles correspond to $1/T=0$ in Eq.~\eqref{eq:inveseT}. We use their masses and widths (as shown in Fig.~\ref{fig:mss_wd}) to obtain two LECs for interactions in the corresponding channels. The solution of the LSE is cutoff dependent, which will be canceled out by the cutoff-dependence of the LECs (practical explicit examples can be seen in Refs.~\cite{Cohen:2004kf,Meng:2020cbk}). Therefore, varying the cutoff
parameter in a reasonable range will not change our results
qualitatively. In this work, we fix the  cutoff scale $\Lambda=1.0$ GeV. We obtain the scattering lengths and effective ranges with
the effective range expansion,
\begin{eqnarray}
T^{-1}(k)=-\frac{\mu}{2\pi}\left(-\frac{1}{a_{s}}-ik+\frac{1}{2}r_{0}k^{2}+...\right).\label{eq:ere}
\end{eqnarray}
 In Fig.~\ref{tab:ere}, we compare our results with
those in Ref.~\cite{Guo:2020vmu}.

\begin{table}
    \centering
    \renewcommand{\arraystretch}{1.5}
    \caption{The scattering lengths $a_s$ and effective ranges $r_0$ extracted from $Z_c$ and $Z_{cs}$ states within the single-channel NREFT (in units of fm). The results in Ref.~\cite{Guo:2020vmu} are listed for comparison.}\label{tab:ere}
    \setlength{\tabcolsep}{0.5mm}
    \begin{tabular}{cccccc}
        \hline \hline
        &  & $Z_{c}(3900)$ & $Z_{c}(4020)$ & $Z_{cs}(3985)$ & $Z_{cs}(4000)$\tabularnewline
        \hline
        & $a_{s}$ & $-0.96\pm0.09$ & $-0.74\pm0.24$ & $-0.76\pm0.26$ & $-0.32\pm0.07$\tabularnewline
        & $r_{0}$ & $-2.88\pm0.37$ & $-3.95\pm2.70$ & $-6.70\pm5.54$ & $-2.08\pm0.80$\tabularnewline
        \hline
        \multirow{2}{*}{\cite{Guo:2020vmu}} & $a_{s}$ & $-0.85\pm0.13$ & $-1.04\pm0.30$ & $-1.00\pm0.74$ & -\tabularnewline
        & $r_{0}$ & $-2.52\pm0.25$ & $-3.90\pm1.35$ & $-4.04\pm1.82$ & -\tabularnewline
        \hline \hline
    \end{tabular}
\end{table}

The $Z_{cs}(3985)$ and $Z_{cs}(4000)$ are the candidates of
$|\bar{\tP}\tV/\bar{\tV}\tP,+\rangle$ and
$|\bar{\tP}\tV/\bar{\tV}\tP,-\rangle$. We do not assign the specific
corresponding relation for now. In general, the
$|\bar{\tP}\tV/\bar{\tV}\tP,+\rangle$ and
$|\bar{\tP}\tV/\bar{\tV}\tP,-\rangle$ states will couple with each
other. The mixing effect arises from the HQSS breaking effect.  We will evaluate the mixing effect in the coupled-channel formalism.

The coupled-channel potential for $Z_{cs}(3985)$ and $Z_{cs}(4000)$
systems can be parameterized as,
\begin{eqnarray}
V(\bm{p},\bm{p}')_{\{\mathtt{\bar{P}V/\bar{V}P},+-\}}
&&=\left[\begin{array}{cc}
    \frac{c_{a}^{+}+\delta c_{a}}{\Lambda} & \frac{\delta c_{a}}{\Lambda}\\
    \frac{\delta c_{a}}{\Lambda} & \frac{c_{a}^{-}+\delta c_{a}}{\Lambda}
\end{array}\right]\nonumber \\
&&+\left[\begin{array}{cc}
    \frac{c_{b}^{+}(\bm{p}^{2}+\bm{p}'^{2})}{\Lambda^{3}}\\
    & \frac{c_{b}^{-}(\bm{p}^{2}+\bm{p}'^{2})}{\Lambda^{3}}
\end{array}\right],\label{eq:cp_ptl}
\end{eqnarray}
where $c_{a/b}^{+/-}$ and $\delta c_a $ are the LECs. The subscripts
$a$ and $b$ represent the LO and NLO, respectively. The superscript
$+/-$ labels the two channels. For the off-diagonal term, we only
keep the LO interaction. For the coupled-channel systems with four
available inputs (masses and widths of two resonances), we have five
unknown LECs to be determined. Thus, a rigorous fitting to determine
the mixing effect is not feasible.

The second best approach is to fix the $c_a^+$,
$c_b^+$, $c_a^-$ and $c_b^-$ from the single-channel calculation and
then vary $\delta c_a$ to check whether the coupled-channel effect
is ignorable or not. The $\delta c_a$ is actually the HQSS breaking
term and its meaning becomes clear when we change the
coupled-channel potential into the $\{\bar{\tP}\tV,\bar{\tV}\tP\}$
basis,
\begin{eqnarray}
    V_{\{\mathtt{\bar{P}V,\bar{V}P}\}}^{1^{+}}=\frac{1}{2\Lambda}\left[\begin{array}{cc}
        c_{a}^{+}+c_{a}^{-} & c_{a}^{+}-c_{a}^{-}\\
        c_{a}^{+}-c_{a}^{-} & c_{a}^{+}+c_{a}^{-}+4\delta c_{a}
    \end{array}\right]+\text{NLO term}.\nonumber\\
\end{eqnarray}
 We also define a ratio $R_{\text{HQSSB}}$ as a reflection of the HQSS breaking effect $R_{\text{HQSSB}}={4\delta c_{a}}/{|c_{a}^{+}+c_{a}^{-}|}.$ The nonzero $R_{\text{HQSSB}}$ indicates  the nonvanishing
off-diagonal terms in Eq.~\eqref{eq:cp_ptl}. Then, the bases
$|\mathtt{\bar{P}V/\bar{V}P},+\rangle$  and
$|\mathtt{\bar{P}V/\bar{V}P},-\rangle$ will mix with each other, in which mixing angle $\theta$ is defined to reflect the
significance of the coupled-channel effect.

\begin{figure}[!htp]
    \centering  \includegraphics[width=0.45\textwidth]{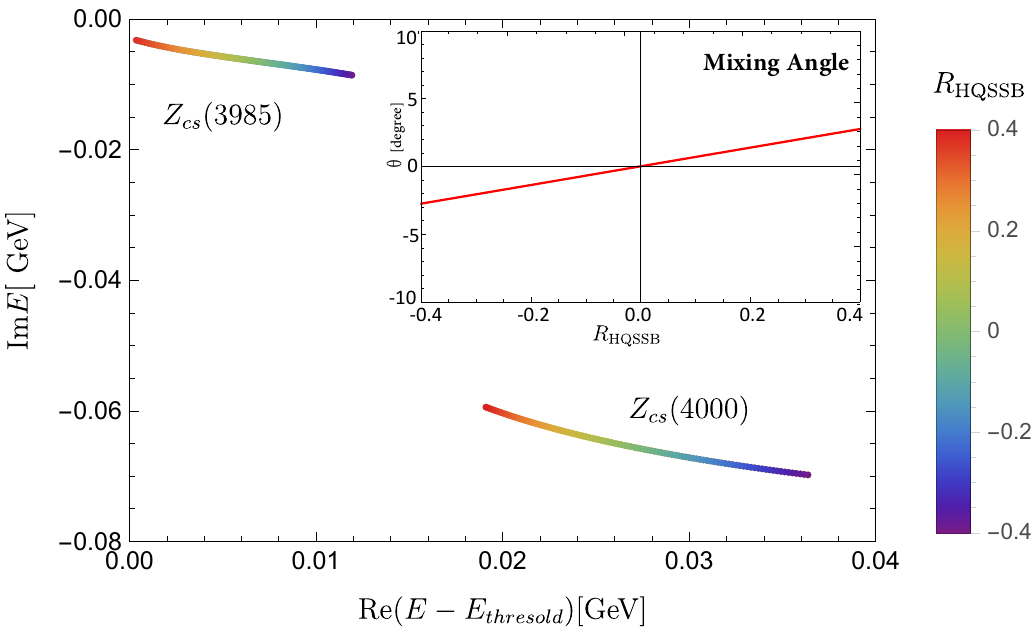}
     \caption{The pole trajectories  and the mixing angle of $|\bar{D}_sD^*/\bar{D}_s^*D,+\rangle$ and $|\bar{D}_sD^*/\bar{D}_s^*D,-\rangle$ for the $Z_{cs}$ states with varying $R_{\text{HQSSB}}$.}\label{fig:traj}
\end{figure}

 We present the pole trajectories and the mixing angle with
varying $R_{\text{HQSSB}}$ in Fig.~\ref{fig:traj}.  When we vary the $R_{\text{HQSSB}}$ from $-0.4$ to
$0.4$, the mixing angle changes quite slightly, less than three degree
and thus negligible. We find that the
40\% HQSS breaking effect does not change the qualitative properties
of two resonances (relative orders for the masses and widths).

The NLO momentum-dependent interaction  in Eq.~\eqref{eq:ptl_single} is very important and necessary  in our framework, which would contribute to the effect range $r_0$ term in ERE and be able to generate a resonance pole above  two-meson threshold even in a single-channel formalism. The $G_0$ in eq.~\eqref{eq:G0} can be expanded as,
	\begin{equation}
		G_0=\frac{\mu}{\pi^{2}}\left(-\Lambda-i\frac{\pi}{2}k+\frac{k^{2}}{\Lambda}+\mathcal{O}(k^{4})\right).
	\end{equation}
If one only includes the LO constant contact potential, from Eqs.~\eqref{eq:Tmatrix}-\eqref{eq:ere}, one can obtain $r_{0}=\frac{4}{\Lambda\pi}$, which is suppressed by the cutoff parameter. With only the LO constant contact term potential, the effective range $r_0$ is small and  only  the bound or virtual states can be produced. An alternative way to make the resonance solution possible in the LO contact interaction is to introduce the coupled channels (See ~\cite{Dong:2020hxe} for a general discussion.). 
 
In principle, one could estimate the LECs in Eq.~\eqref{eq:ptl_single} through the meson-exchange model.  For $Z_{cs}$ states, it was shown that the exchange of ground light vector mesons is not allowed~\cite{Aceti:2014kja,Aceti:2014uea,Ikeno:2021ptx}. In contrast, the exchange of pseudoscalar $\eta$ meson could be introduced from the chiral Lagrangian~\cite{Wang:2020htx}. If we follow the path of Bonn meson-exchange model~\cite{Machleidt:1987hj}, the scalar-meson-exchange~\cite{Chen:2020yvq} and two-kaon-exchange~\cite{Wang:2020htx} interactions also contribute to the $\bar{D}_s^*D/\bar{D}_sD^*$ potential. Moreover, the heavy-meson-exchange interaction in Ref.~\cite{Ikeno:2021ptx} might also have contribution. However, in practice, estimating the contact LECs from the short-range part of the meson-exchange interaction is very involved. First, some of the coupling constants in the meson-exchange model for $\bar{D}_s^*D/\bar{D}_sD^*$ are not available. Meanwhile, in Ref.~\cite{Epelbaum:2003xx}, it was shown that the contact interaction of EFT agrees with short-range part of meson-exchange interaction only when the EFT and the meson-exchange model were calculated to a high precision. Thus, it is hard to give a very reasonable estimation of the LECs through a low order (NLO) calculation from a rough meson-exchange model.

In Refs.~\cite{Ikeno:2021ptx,Aceti:2014kja,Aceti:2014uea}, the authors adopted a coupled-channel formalism to investigate $Z_c$ and $Z_{cs}$ states. It was found that though the $Z_{cs}$ state couples to the $J/\psi K$ and $\eta_cK^*$ channels weakly, the two channels play important roles~\cite{Ikeno:2021ptx}. In Ref~\cite{Meng:2020ihj}, we concluded differently that the  $J/\psi \pi (K)$ channel neither changes the property of the pole nor contributes to the large partial decay width. From our perspective,  the interacting mechanism and physical interpretation of $Z_{cs}$ in Ref.~\cite{Ikeno:2021ptx} are different with those in this manuscript and Ref~\cite{Meng:2020ihj}.  In Ref.~\cite{Ikeno:2021ptx}, the interactions are determined  within the local hidden gauge approach in SU(4) flavor symmetry through heavy vector meson exchange. The diagonal potentials are small, then the off-diagonal potentials arising from the coupled-channel effect with the $J/\psi \pi$, etc., becomes important. The authors explained the $Z_{cs}$ as a cusp effect. In our work, we presume $Z_c$ and $Z_{cs}$ are resonances considering their masses are above the related thresholds.  We choose to determine the low energy constants by fitting the experimental data. Given different premises, it is not strange to obtain different significance of the coupled-channel effect in two framework.

\section{Phenomenological analysis and conclusion}
We combine the experimental results
~\cite{Zyla:2020zbs,Ablikim:2020hsk,Aaij:2021ivw} and our
calculations to analyze the implications of
$Z_{cs}(3985)$ and $Z_{cs}(4000)$ as two different states. The main
conclusions are illustrated in Fig.~\ref{fig:twocase} (b) and explained
as follows,
\begin{enumerate}
\item {\it HQSS is a good approximation for the $\bar{D}_s^{(*)}D^{(*)}$ systems.}
The consistent properties of $Z_{c}(3900)$ and $Z_c(4020)$ indicate that HQSS works
well for  $\bar{D}^{(*)}D^{(*)}$ systems. Meanwhile, the HQSS violation effect
for the $|\bar{D}_sD^*/\bar{D}_s^*D,+\rangle$ and $|\bar{D}_sD^*/\bar{D}_s^*D,-\rangle$
systems is negligible. In Fig.~\ref{fig:traj}, the $40\%$ HQSS breaking effect will neither
change the qualitative properties of two resonances,
nor give rise to considerable mixing effect.

\item {\it $Z_{cs}(4000)$ and $Z_{cs}(4220)$ are the HQSS partners.}
$Z_{cs}(4220)$ is a wide resonance above the $\bar{D}_s^*D^*$
threshold. It is natural to interpret it as $\bar{D}_s^*D^*$ resonances
with $J^P=1^+$. Its HQSS partner state $|\bar{D}_s^*D/\bar{D}_sD^*,+\rangle$ should be a wide resonance as $Z_{cs}(4220)$
because of the same interactions in the heavy quark limit in Eq.~\eqref{eq:hqss_vector}.  Thus, it
is reasonable to infer that the wider resonance $Z_{cs}(4000)$
rather than the narrower resonance $Z_{cs}(3985)$ is the partner of
the $Z_{cs}(4220)$ state.

\item {\it $Z_{cs}(4000)$ and $Z_{cs}(3985)$ are almost pure
$|\bar{D}_s^*D/\bar{D}_sD^*,+\rangle$ and $|\bar{D}_s^*D/\bar{D}_sD^*,-\rangle$ states,
respectively.} As the HQSS partner of $Z_{cs}(4220)$,
the hadronic component of the $Z_{cs}(4000)$ is $|\bar{D}_s^*D/\bar{D}_sD^*,+\rangle$. If the $Z_{cs}(3985)$ is the second state near the $\bar{D}_s^*D/\bar{D}_sD^*$ threshold, it should be dominated by the $|\bar{D}_s^*D/\bar{D}_sD^*,-\rangle$ component. In Fig.~\ref{fig:traj}, the mixing effect of two different components is very tiny.

\item {\it There should exist a tensor $\bar{D}_s^*D^*$ state as the HQSS partner of $Z_{cs}(3985)$. }
As illustrated in  Eq.~\eqref{eq:hqss_tensor}, the interaction in
the tensor state $\langle
V_{q\bar{q}}^{s}\rangle_{\{\bar{\mathtt{V}}\mathtt{V}\}}^{2^{+}}$ is
the same as that of $Z_{cs}(3985)$. With the interaction extracted
from the $Z_{cs}(3985)$, we give a prediction of the mass and decay
width of the tensor state,
    \begin{equation}
        M=4126\pm3~\text{ MeV},\quad\Gamma=13\pm6~\text{ MeV}.
    \end{equation}

\item {\it The branch ratio $\mathcal{R}(Z_{cs}\to \bar{D}_s^*D/Z_{cs}\to \bar{D}_sD^*)\approx 0.5$ for $Z_{cs}(3985)$  or  $Z_{cs}(4000)$ states.} It is the consequence of that $Z_{cs}(4000)$ and $Z_{cs}(3985)$ are almost pure $|\bar{D}_s^*D/\bar{D}_sD^*,+\rangle$ and $|\bar{D}_s^*D/\bar{D}_sD^*,-\rangle$ states, respectively.

\item {\it The decay mode $Z_{cs}(3985)\to J/\psi K$ is suppressed in the HQSS limit.}
In Eqs.~\eqref{eq:rearange1}-\eqref{eq:rearange4}, we decompose the
$Z_{cs}(3985)$, $Z_{cs}(4000)$, $Z_{cs}(4220)$ and our prediction
$Z_{cs}(4126)$ into the light and heavy degrees of freedom, which
are separately conserved in the heavy quark limit. For example, the flavor-spin wave function of $|\bar{D}^*D^*,2^+\rangle$ is $ |1^-_{\bar{c}q_1},1^-_{c\bar{q}_2};2^+\rangle$. In the heavy quark spin symmetry, we have $ |1^-_{\bar{c}q_1},1^-_{c\bar{q}_2};2^+\rangle=|1_{\bar{c}c}^{-},1_{q_{1}\bar{q}_{2}}^{-};2^{+}\rangle.$ In Fig.~\ref{fig:twocase} (b), we use $|J/\psi\rho\rangle$ with total spin $J=2$ to label the wave function after rearrangement. This leads to
a series of selection rules~\cite{Braaten:2014qka,Ma:2014zva,Wang:2015qlf} as
displayed in Fig.~\ref{fig:twocase} (b). For instance, the decay process
$Z_{cs}(3985)\to J/\psi K$ is forbidden within the HQSS. Another way to derive these selection rules is the
conservation of the $G_{I/U/V}$  parity. For example, the $G_{U/V}$
parities of $Z_{cs}(3985)$, $J/\psi$ and $K$ are $-1$, $-1$ and
$-1$, respectively.  The decay $Z_{cs}(3985)\to J/\psi K$ is
suppressed by the  $G_{U/V}$ parity conservation. The
$G_{I/U/V}$-parity encodes the spin information into its charge
conjugation part. Thus, it is natural to obtain the same results from the
HQSS decomposition and $G_{I/U/V}$ analysis. As illustrated in Fig.~\ref{fig:twocase} (b), the $Z_{cs}(3985)$ state is more likely to be
observed in the $\bar{D}_s^*D/\bar{D}_sD^*$ channel, while it is
hard to be observed in the $J/\psi K$ channel, which is consistent
with the experimental results~\cite{Ablikim:2020hsk}. Similar
results have also been obtained in Ref.~\cite{Chen:2021erj}.
Finally, one should note that the above selection rules of the
$G_{I/U/V}$ parity may be violated by the $\SU3$ flavor symmetry and
heavy quark symmetry breaking effects.

\item {\it The violation effect of the $\SU3$ symmetry might be significant.}
As shown in Fig.~\ref{fig:twocase}, the widths of the $Z_c(3900)$ and
$Z_{c}(4020)$ states are much smaller than those of their $\SU3$
partners $Z_{cs}(4000)$ and $Z_{cs}(4220)$. Moreover, the
$I^{G_I}(J^{PC})=1^-(1^{++})$ state, as the partner state of
$Z_{cs}(3985)$ in the $\SU3$ limit, is missing in experiments.
Therefore, the assignment of the $Z_{cs}(3985)$ and $Z_{cs}(4000)$
as two different states implies the large
$\SU3$ breaking effect.
\end{enumerate}

The implications of $Z_{cs}(3985)$ and $Z_{cs}(4000)$ in
Fig.~\ref{fig:twocase} (b) are quite different from the consequences of
there existing only one $[\bar{D}_sD^*/\bar{D}_s^*D]^{1^+}$ state in
Fig.~\ref{fig:twocase} (a). In the theoretical aspects, the
$Z_{cs}(3985)$ and $Z_{cs}(4000)$ as two different states implies that
the $Z_{cs}(3985)$ is the $|\bar{D}_s^*D/\bar{D}_sD^*,-\rangle$
state, which is different from the assignment as the
$|\bar{D}_s^*D/\bar{D}_sD^*,+\rangle$ in Fig.~\ref{fig:twocase} (a). Meanwhile, the significant $\SU3$ violation implied by
$Z_{cs}(3985)$ and $Z_{cs}(4000)$ as two states contradicts the
presuming $\SU3$ symmetry in Fig.~\ref{fig:twocase} (a).

 With $Z_{cs}(3985)$ and $Z_{cs}(4000)$ as two different states, the
 decay $Z_{cs}(3985)\rightarrow J/\psi K$ is forbidden in the heavy
 quark limit. Meanwhile, an additional tensor state $\bar{D}_s^*D^*$ is
predicted as the HQSS partner of the $Z_{cs}(4000)$ in the two-state
framework. If one assumes that there only exists one
$[\bar{D}_sD^*/\bar{D}_s^*D]^{1^+}$ state as shown in
Fig.~\ref{fig:twocase} (a), no tensor partner is predicted and the
$Z_{cs}(3985) \to J/\psi K$ decay mode is allowed in the HQSS limit.
 In the
experimental aspects, searching for the tensor $\bar{D}_s^*D^*$
state and the $Z_{cs}(3985) \to J/\psi K$ decay mode can  be used to distinguish the
interpretation of the $Z_{cs}(3985)$ and $Z_{cs}(4000)$ as two
different states from there only existing one
$\bar{D}_s^*D/\bar{D}_sD^*$ resonance.

\vspace{0.5cm}

\begin{acknowledgements}
We are grateful to the helpful discussions with Prof.~Makoto Oka  and Xin-Zhen Weng. This
project was supported by the National Natural Science Foundation of
China (11975033 and 12070131001). This project was also funded by
the Deutsche Forschungsgemeinschaft (DFG, German Research
Foundation, Project ID 196253076-TRR 110). G.J. Wang was supported
by JSPS KAKENHI (No.20F20026).
\end{acknowledgements}

\bibliography{twozcc_sc2.6}

\end{document}